# NEGATIVE THERMAL EXPANSION OF $ZrW_2O_8$ AND $HfW_2O_8$: CONTRADICTION TO A THERMODYNAMIC RELATION

## I.A. STEPANOV[1]

*Institute of Physical Chemistry II, Albertstrasse 23a, Freiburg 79104, Germany*

It has been shown that the dependence of negative thermal expansion coefficient of $ZrW_2O_8$ and $HfW_2O_8$ on the temperature found by many authors contradicts to an important thermodynamic relation $(\partial C_P/\partial P)_T = -TV((\partial\alpha/\partial T)_P + \alpha^2)$.
**Keywords:** negative thermal expansion, heat capacity, $ZrW_2O_8$, $HfW_2O_8$, the 1st law of thermodynamics

In [1-9], dependence of the volume on the temperature and heat capacity $C_P$ on the temperature has been obtained for $ZrW_2O_8$ and $HfW_2O_8$. Both substances have cubic crystal structure. There is a thermodynamic relation using which one can verify the 1st law of thermodynamics [10, 11]:

$$(\partial C_P/\partial P)_T = -TV((\partial\alpha/\partial T)_P + \alpha^2) \tag{1}$$

where $\alpha$ is the thermal expansion coefficient. One can show that the dependence $\alpha(T, P)$ found in [1-6] contradicts to the relation (1) for negative $\alpha$.

In [1-6], for both substances, $(\partial\alpha/\partial T)_P < 0$ in big temperature intervals, because $V \approx V_0 - bT$, $b = \text{const}$. If $V \approx V_0 - bT$ than $(\partial\alpha/\partial T)_P + \alpha^2 \approx -b^3 T/V_0^3$. $(\partial C_P/\partial T)_P > 0$ in the same temperature intervals. $C_P$ increases with the temperature from $T \approx 300$ K till $T = T_{max}$ ($T_{max} \approx 440$ K for $ZrW_2O_8$), then it decreases (till $T \approx 470$ K for $ZrW_2O_8$) K, then it again increases (at $T > 470$ K for $ZrW_2O_8$) [2, 7], *Figure 1*. In [8, 9] $C_P(T)$ has been measured from 0 till 300 K. It increases drastically from $T = 0$ K till $T \approx 100$ K, *Figure* 2.

In TABLES 1- 3 and *Figures* 3 and 4 dependence of the lattice parameter and $\alpha$ of $ZrW_2O_8$ and $HfW_2O_8$ on the temperature is given. The data from TABLE 1 are from [3], they were provided by Prof. Evans, the data from TABLE 2 are from [4], they were provided by Prof. Sleight, the data from TABLE 3 are from [2], they were provided by Prof. Tsuji, *Figure 4* was provided by Prof. Hashimoto. The data in [5] coincide with the data from [1] within the error limits in the interval $300 < T < 420$ K. The dependence $\alpha(T)$ is found using numerical differentiation of the data.

---

[1] *Permanent address: Institute of Chemical Physics, Latvian University, Rainis bulv. 19, Riga, LV-1586, Riga, Latvia*

From TABLES 1 and 2 and *Figures* 3 and 4 one can deduce that at 0<T<100 K α of $ZrW_2O_8$ steeply decreases. From T≈400 K till T=$T_{min}$≈440 K α decreases drastically, then it steeply increases till T≈470 K, then it again decreases. There $|(\partial\alpha/\partial T)_P|>>\alpha^2$.

The data for $HfW_2O_8$ in TABLE 3 are insufficient to built a smooth curve but a drastic minimum at T≈460 K is evident. The dependence $C_P(T)$ has a drastic maximum at this point [2], *Figure* 1.

However, one can prove that

sign $(\partial C_P/\partial P)_T$ = sign $(dC_P/dP)$ = sign $(dC_P/dT)(dT/dP)$.                              (2)

$C_P = C_P(P, T)$. A special case is $C_P=C_P(T(P))$. $dC_P/dP$ has a certain sign at T≠const in an infinitively small vicinity of the point T=const. At the point T=const it will have the same sign due to continuity of the function $C_P(P, T)$.

Let us consider the temperature interval 0<T<100 K for $ZrW_2O_8$. For α<0, dT/dP<0 and $dC_P/dT$>0 whence the left part of (1) $(\partial C_P/\partial P)_T$<0. The right side of (1) has positive sign. It is a contradiction. In the temperature interval 400<T<440 K for $ZrW_2O_8$ $(\partial C_P/\partial P)_T$<0 but the right side of (1) is positive. In the temperature interval 440<T<470 K $(\partial C_P/\partial P)_T$>0 but the right side of (1) is negative. At T> 470 K, $(\partial C_P/\partial P)_T$<0 but $-(\partial\alpha/\partial T)_P$>0.

A possible explanation of this paradox can be found in [11]. There it has been supposed that for substances with negative thermal expansion the 1[st] law of thermodynamics must have the following form: dQ=dU-PdV. If to derive (1) using this formula, one obtains

$(\partial C_P/\partial P)_T = TV((\partial\alpha/\partial T)_P+\alpha^2)$.                                                      (3)

TABLE 1. Dependence of the lattice constant of $ZrW_2O_8$ on the temperature [3]

| T, K | a, Å | $\alpha = 1/V(\partial V/\partial T)_P, *10^4, K^{-1}$ |
|---|---|---|
| 2 | 9.1800 | 0.0395 |
| 20 | 9.1797 | -0.1367 |
| 40 | 9.1785 | -0.2364 |
| 60 | 9.1769 | -0.2903 |
| 80 | 9.1750 | -0.3186 |
| 100 | 9.1731 | -0.2991 |
| 120 | 9.1713 | -0.3001 |
| 140 | 9.1694 | -0.3159 |
| 160 | 9.1675 | -0.3012 |
| 180 | 9.1657 | -0.2955 |
| 200 | 9.1639 | -0.2842 |
| 220 | 9.1622 | -0.2864 |
| 240 | 9.1604 | -0.2894 |
| 260 | 9.1587 | -0.2755 |
| 280 | 9.1570 | -0.2791 |
| 300 | 9.1553 | -0.2789 |
| 320 | 9.1536 | -0.2766 |
| 340 | 9.1519 | -0.2862 |
| 360 | 9.1501 | -0.2995 |
| 380 | 9.1484 | -0.2372 |
| 400 | 9.1470 | -0.2767 |
| 420 | 9.1447 | -0.4762 |
| 440 | 9.1432 | -0.4752 |
| 460 | 9.1416 | -0.6746 |
| 480 | 9.1394 | -0.5675 |
| 500 | 9.1384 | -0.2066 |
| 520 | 9.1372 | -0.1859 |

TABLE 2. Dependence of the lattice constant of $ZrW_2O_8$ on the temperature [4]

| T, K | a, Å | $\alpha = 1/V(\partial V/\partial T)_P$, $*10^4$, $K^{-1}$ |
|---|---|---|
| 2 | 9.1800 | 0.0577 |
| 22 | 9.1796 | -0.1655 |
| 42 | 9.1783 | -0.2534 |
| 62 | 9.1767 | -0.2770 |
| 82 | 9.1749 | -0.3059 |
| 102 | 9.1730 | -0.3042 |
| 122 | 9.1711 | -0.3071 |
| 142 | 9.1692 | -0.3172 |
| 162 | 9.1673 | -0.3086 |
| 182 | 9.1654 | -0.2893 |
| 202 | 9.1637 | -0.2871 |
| 222 | 9.1619 | -0.2859 |
| 242 | 9.1602 | -0.2784 |
| 262 | 9.1585 | -0.2806 |
| 282 | 9.1568 | -0.2797 |
| 302 | 9.1551 | -0.2764 |
| 322 | 9.1534 | -0.2811 |
| 342 | 9.1517 | -0.2706 |
| 362 | 9.1501 | -0.2986 |
| 382 | 9.1482 | -0.2561 |
| 402 | 9.1468 | -0.2902 |
| 422 | 9.1444 | -0.4673 |
| 442 | 9.1412 | -0.5572 |
| 462 | 9.1382 | -0.3410 |
| 482 | 9.1371 | -0.1370 |
| 502 | 9.1360 | -0.1945 |
| 520 | 9.13489 | -0.2204 |

TABLE 3. Dependence of the lattice constant of $HfW_2O_8$ on the temperature [2]

| T, K | a, Å | $\alpha = 1/V(\partial V/\partial T)_P$, *$10^4$, $K^{-1}$ |
|---|---|---|
| 91 | 9.1469 | -0.3075 |
| 121 | 9.1441 | -0.3056 |
| 171 | 9.1396 | -0.2867 |
| 220 | 9.1357 | -0.2158 |
| 270 | 9.1325 | -0.2669 |
| 299 | 9.1300 | -0.2390 |
| 319 | 9.1286 | -0.2904 |
| 338 | 9.1266 | -0.3184 |
| 358 | 9.1251 | -0.2096 |
| 377 | 9.1238 | -0.2665 |
| 396 | 9.1221 | -0.2549 |
| 415 | 9.1209 | -0.1970 |
| 435 | 9.1194 | -0.3440 |
| 454 | 9.1168 | -0.5422 |
| 473 | 9.1138 | -0.3853 |
| 501 | 9.1122 | -0.1467 |
| 520 | 9.1111 | -0.2010 |
| 539 | 9.1101 | -0.1514 |
| 558 | 9.1094 | -0.0894 |

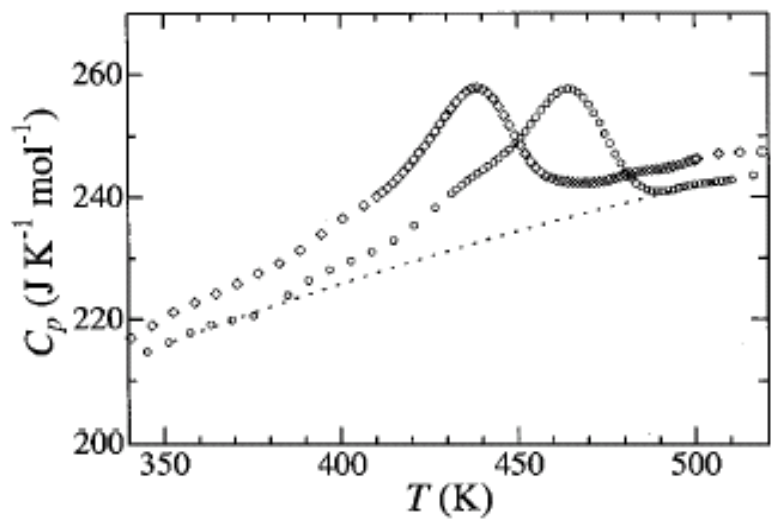

*Figure 1*. Heat capacities of HfW$_2$O$_8$ (○) and ZrW$_2$O$_8$ (◊) as a function of temperature. The broken line is a baseline used to estimate the excess heat capacity for HfW$_2$O$_8$ [2].

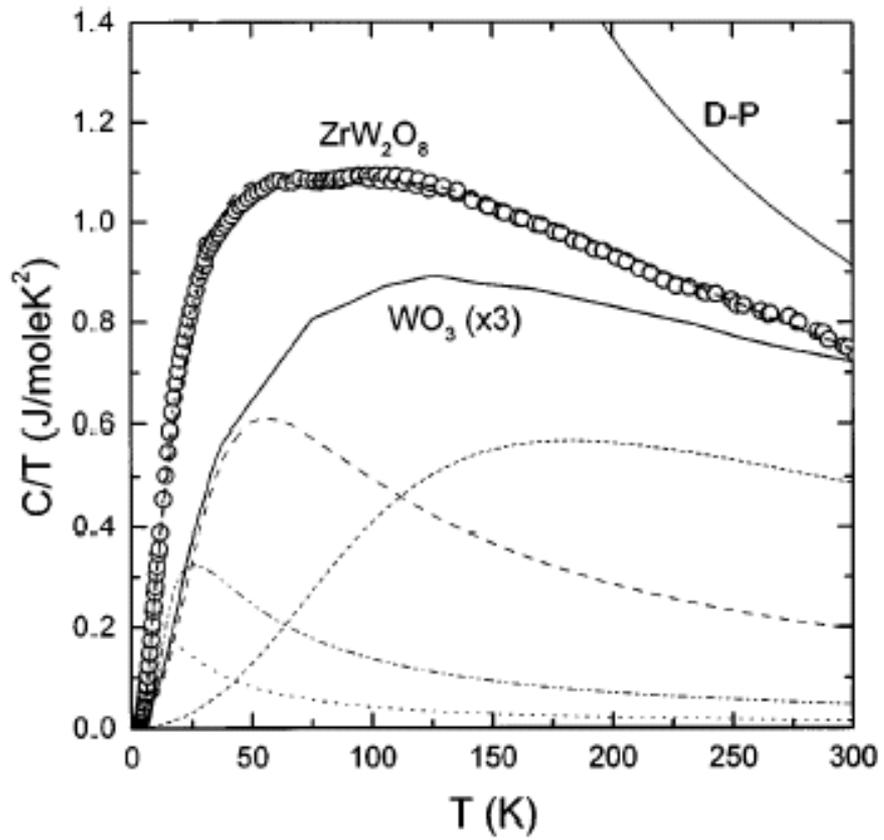

*Figure 2*. The specific heat divided by temperature, C(T)/T for $ZrW_2O_8$. The dashed and dotted curves show the individual contributions of the fitting terms. The solid line is C(T) for $WO_3$, multiplied by three to facilitate comparison with $ZrW_2O_8$. Also shown is the Dulong-Petit (D-P) specific heat for $ZrW_2O_8$ [9].

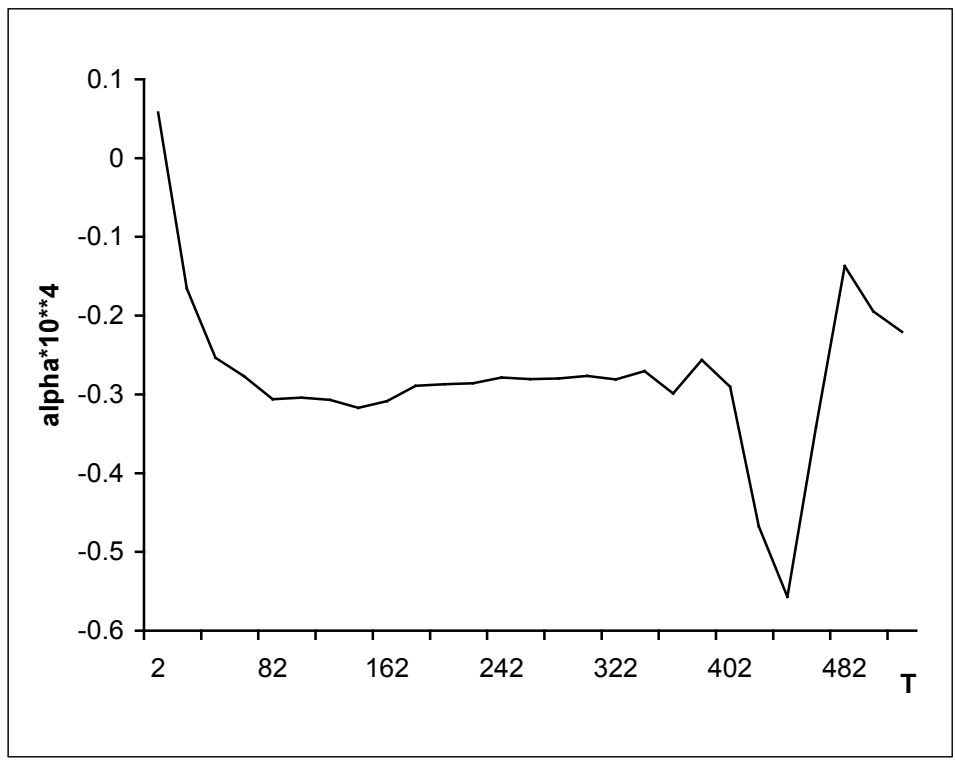

*Figure. 3*. Dependence of the thermal expansion coefficient of $ZrW_2O_8$ on the temperature (TABLE 2).

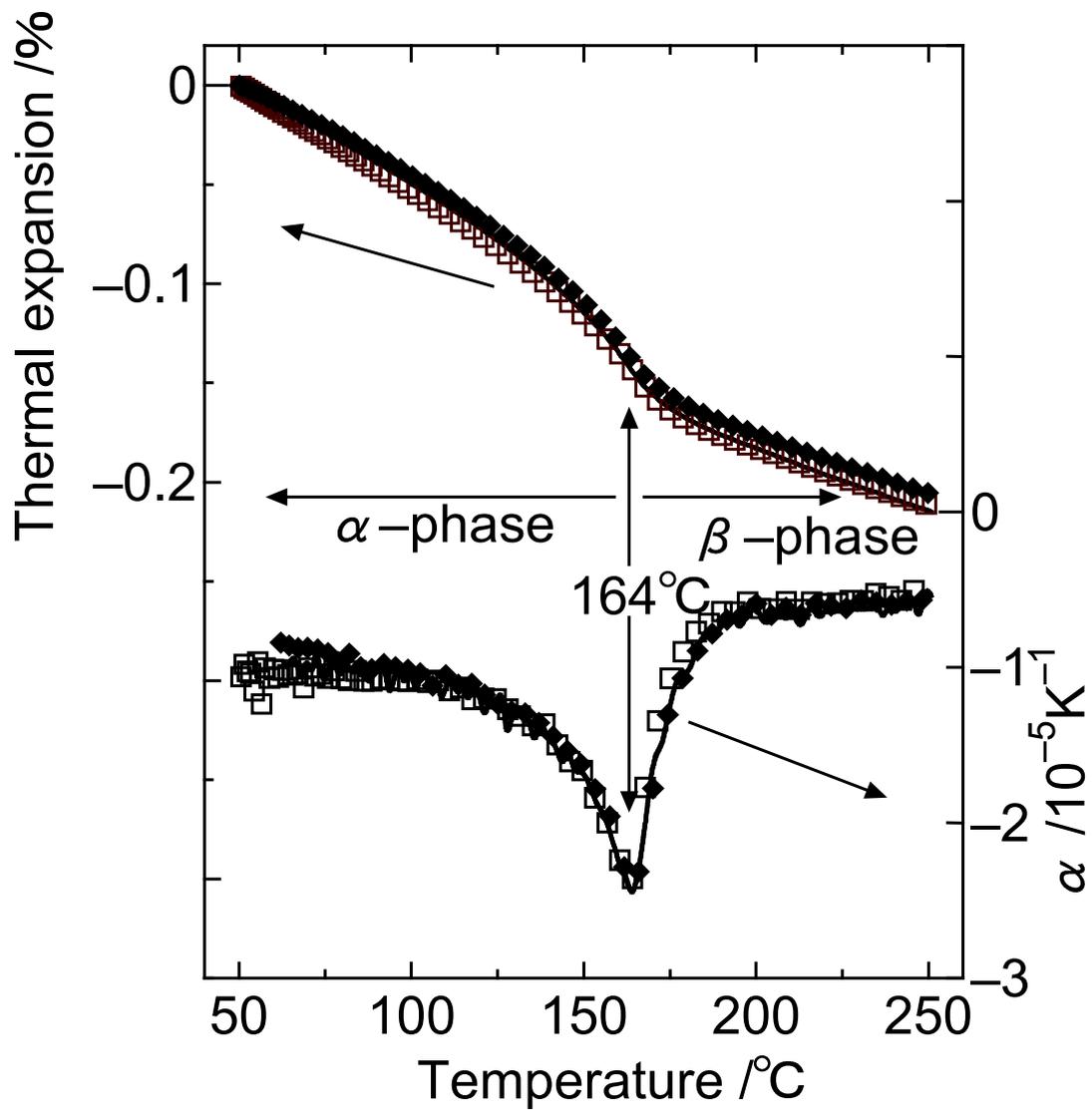

*Figure 4.* Thermal expansion and thermal expansion coefficient, α, of $ZrW_2O_8$ ceramic specimen measured with dilatometer. A λ-type transition was observed at 164 °C. Stable $ZrW_2O_8$ phase is depicted in the *Figure* [6].